\documentclass[10pt,twocolumn,letterpaper]{article}

\usepackage{iccv}
\usepackage{times}
\usepackage{epsfig}
\usepackage{graphicx}
\usepackage{amsmath}
\usepackage{amssymb}

\usepackage{subfigure}
\usepackage{multirow}
\usepackage{verbatim}
\usepackage{booktabs}
\usepackage{pifont}
\usepackage{tablefootnote}

\usepackage[breaklinks=true,bookmarks=false]{hyperref}

\newcommand{\tabincell}[2]{\begin{tabular}{@{}#1@{}}#2\end{tabular}}

\iccvfinalcopy 

\def\iccvPaperID{5576} 

\ificcvfinal\fi

\begin{document}

\title{Non-Local ConvLSTM for Video Compression Artifact Reduction}

\author{Yi Xu\textsuperscript{1}\thanks{This author did most work during his internship at Bilibili.}\qquad
Longwen Gao\textsuperscript{2}\qquad
Kai Tian\textsuperscript{1}\qquad
Shuigeng Zhou\textsuperscript{1}\thanks{Corresponding author.}\qquad
Huyang Sun\textsuperscript{2}\\
\textsuperscript{1}Shanghai Key Lab of Intelligent Information Processing, and School of \\Computer Science, Fudan University, Shanghai, China\\
\textsuperscript{2}Bilibili, Shanghai, China\\
{\tt\small \{yxu17,ktian14,sgzhou\}@fudan.edu.cn} \quad
{\tt\small \{gaolongwen,sunhuyang\}@bilibili.com}
}

\maketitle
\thispagestyle{empty}

\begin{abstract}
Video compression artifact reduction aims to recover high-quality videos
from low-quality compressed videos. Most existing approaches use a single neighboring frame or a pair of neighboring frames (preceding and/or following the target frame) for this task. Furthermore, as frames of high quality overall may contain low-quality patches, and high-quality patches may exist in frames of low quality overall, current methods focusing on nearby peak-quality frames~(PQFs) may miss high-quality details in low-quality frames. To remedy these shortcomings, in this paper we propose a novel end-to-end deep neural network called non-local ConvLSTM (NL-ConvLSTM in short) that exploits multiple consecutive frames. An approximate non-local strategy is introduced in NL-ConvLSTM to capture global motion patterns and trace the spatiotemporal dependency in a video sequence. This approximate strategy makes the non-local module work
in a fast and low space-cost way. Our method uses the preceding and following frames of the target frame to generate a residual, from which a higher quality frame is reconstructed. Experiments on two datasets show that NL-ConvLSTM outperforms the existing methods. \end{abstract}

\section{Introduction}
Video compression algorithms are widely used due to limited communication bandwidth and storage space in many real (especially mobile) application senarios~\cite{sullivan2012overview}. While significantly reducing the cost of transmission and storage, lossy video compression also leads various compression artifacts such as blocking, edge/texture floating, mosquito noise and jerkiness~\cite{zeng2014characterizing}.
Such visual distortions often severely impact the quality of experience~(QoE).
Consequently, video compression artifact reduction has emerged as an important research topic in multimedia and computer vision areas~\cite{lu2018deep,yang2018enhancing, yang2018multi}.

\begin{figure}[t]
	\centering
	\includegraphics[width=\linewidth]{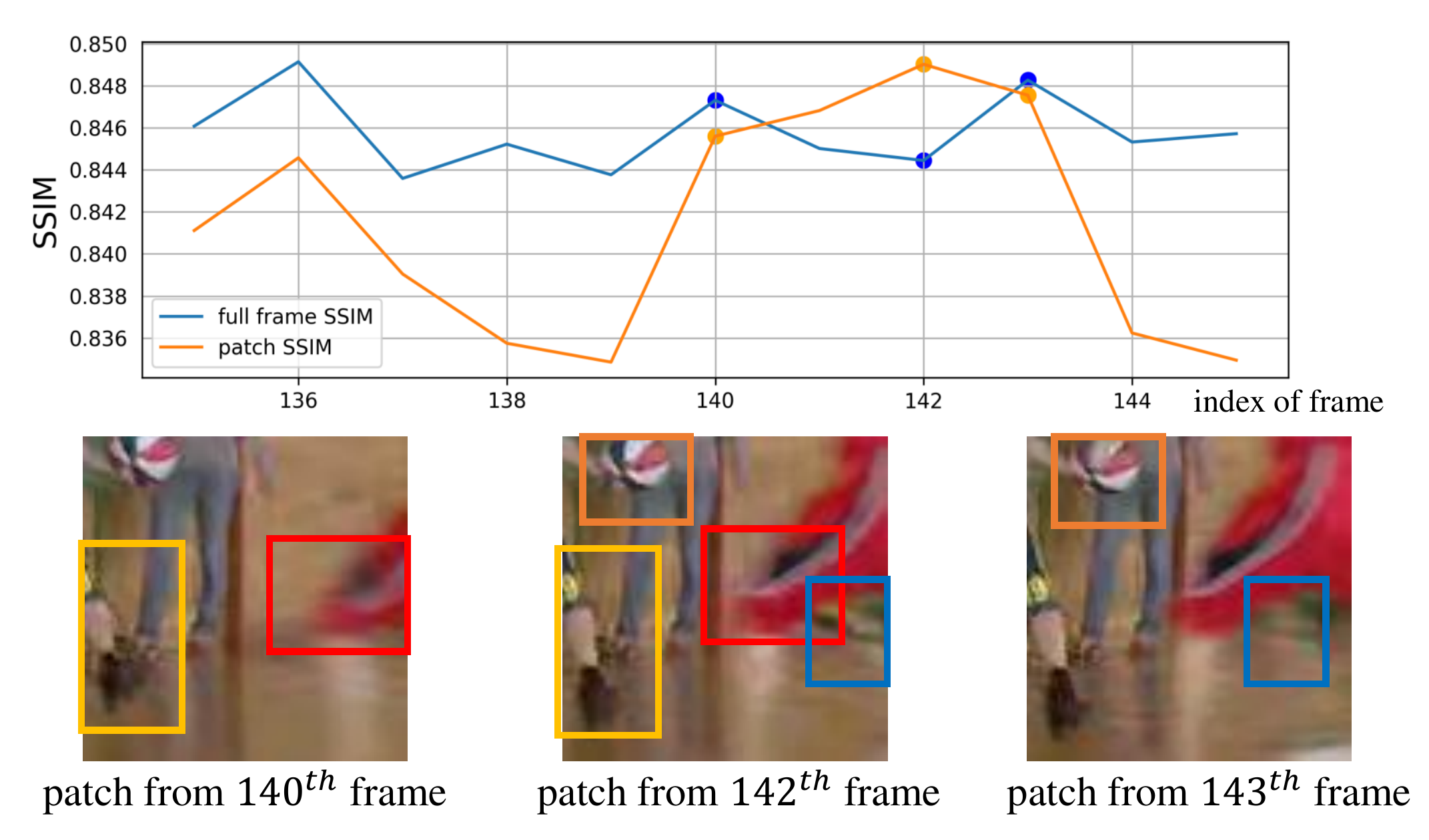}
	
	\caption{An example of high-quality patches existing in low-quality frames. Here, though the $140^{th}$ and $143^{th}$ frames have better full-frame SSIM than the $142^{th}$ frame, the cropped patch from the $142^{th}$ frame has the best patch SSIM. The upper part shows the SSIM values of the fames and cropped patches; the lower images are the cropped patches from three frames. Also, comparing the details in the boxes of the same color, we can see that the cropped patches from the $142^{th}$ frame are of better quality.
	}
	\vspace{-5pt}
	\label{fig:quality varing frames}
\end{figure}

In recent years, significant advances in compressed image/video enhancement have been achieved due to the successful applications of deep neural networks.
For example, \cite{dong2015compression,galteri2017deep,tai2017memnet,zhang2017beyond} directly utilize deep convolutional neural networks to remove compression artifacts of images without considering the characteristics of underlying compression algorithms. \cite{jin2018quality,wang2017novel,yang2018enhancing,yang2017decoder} propose models that are fed with compressed frames and output the enhanced ones. These models all use a single frame as input, do not consider the temporal dependency of neighboring frames. To exploit the temporal correlation of neighboring frames, \cite{lu2018deep} proposes the deep kalman filter network, \cite{xue2019video} employs task oriented motions, and \cite{yang2018multi} uses two motion-compensated nearest PQFs. However, \cite{lu2018deep} uses only the preceding frames of the target frame,
while \cite{xue2019video,yang2018multi} adopt only a pair of neighboring frames, which may miss high-quality details of some other neighbor frames (will be explained later).

Video compression algorithms have intra-frame and inter-frame codecs. Inter coded frames (P and B frames) 
significantly depend on the preceding and following neighbor frames. Therefore, extracting spatiotemporal
relationships among the neighboring frames can provide useful information for improving video enhancement performance.
However, mining detailed information from one/two neighboring frame(s) or even two nearest PQFs is not enough for compression video artifact reduction.
To illustrate this point, we present an example in Fig.~\ref{fig:quality varing frames}. Frames with a larger \emph{structural similarity index measure}~(SSIM) are usually regarded as better visual quality. Here, though the $140^{th}$ and $143^{th}$ frames have better overall visual quality than the $142^{th}$ frame, the cropped patch of highest quality comes from the $142^{th}$ frame. The high-quality details in such patches would be ignored if mining spatiotemporal information from videos using the existing methods.

Motivated by the observation above, in this paper we try to capture the hidden spatiotemporal information from multiple preceding and following frames of the target frame for boosting the performance of video compression artifact reduction. To this end, we develop a non-local ConvLSTM framework that uses the non-local mechanism~\cite{buades2005non} and ConvLSTM~\cite{xingjian2015convolutional} architecture to learn the spatiotemporal information from a frame sequence.
To speed up the non-local module, we further design an approximate and effective method to compute the inter-frame pixel-wise similarity. Comparing with the existing methods, our method is advantageous in at least three aspects: 1) No accurate motion estimation and compensation is explicitly needed;
2) It is applicable to videos compressed by various commonly-used compression algorithms such as H.264/AVC and H.265/HEVC;
3) The proposed method outperforms the existing methods.

Major contributions of this work include:
1) We propose a new idea for video compression artifact reduction by exploiting multiple preceding and following frames of the target frame, without explicitly computing and compensating motion between frames. 2) We develop an end-to-end deep neural network called non-local ConvLSTM to learn the spatiotemporal information from multiple neighboring frames. 
3) We design an approximate method to compute the inter-frame pixel-wise similarity, which dramatically reduces calculation and memory cost. 
4) We conduct extensive experiments over two datasets to evaluate the proposed method, which achieves state-of-the-art performance for video compression artifact reduction.
\begin{figure*}
	\begin{center}
		\includegraphics[width=\linewidth]{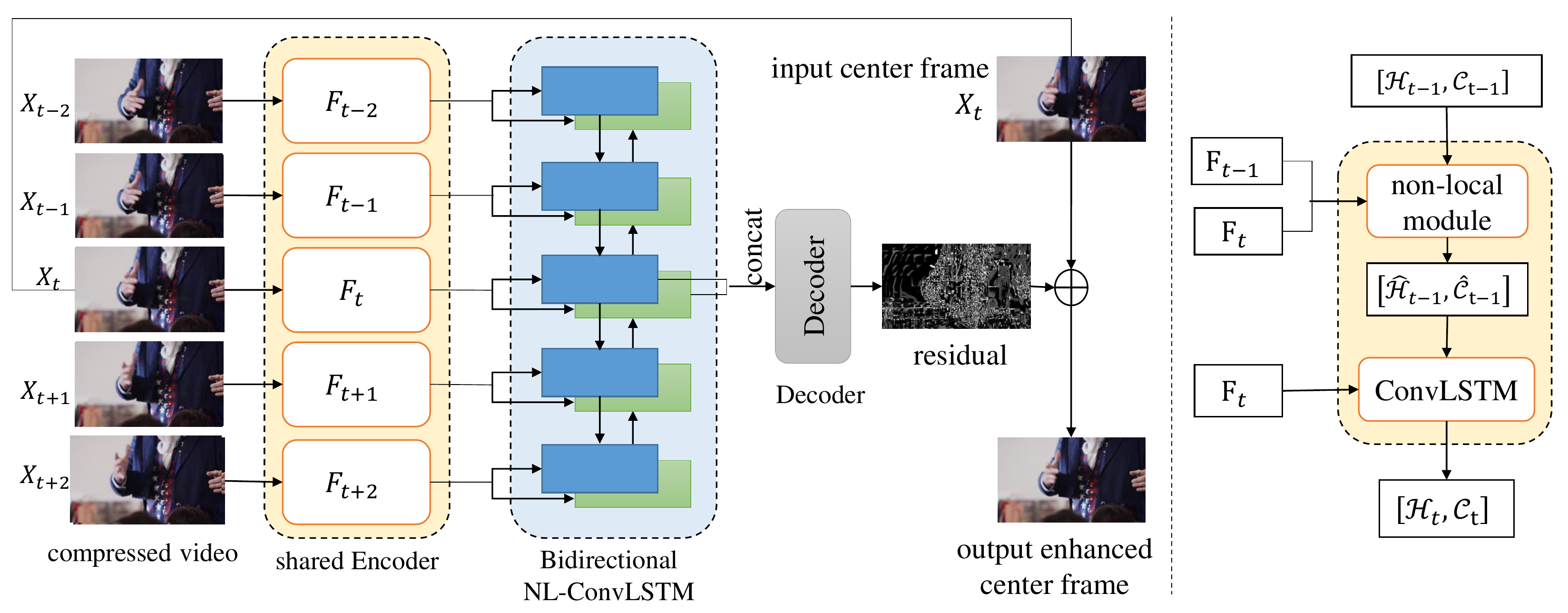}
	\end{center}
	\vspace{-5pt}
	\caption{The framework of our method~(left) and the architecture of NL-ConvLSTM~(right)}
	\label{fig:architecture}
\end{figure*}

\section{Related Work}
\subsection{Single Image Compression Artifact Reduction}

Early works mainly include manually-designed filters~\cite{buades2005non,dabov2007image,ramamurthi1986nonlinear,zhang2013compression}, iterative approaches based on the theory of projections onto convex sets~\cite{minami1995optimization,zakhor1992iterative}, wavelet-based methods~\cite{wu2001efficient} and sparse coding~\cite{chang2014reducing,liu2015data}.

Following the success of AlexNet~\cite{krizhevsky2012imagenet} on ImageNet~\cite{russakovsky2015imagenet}, many deep learning based methods have been applied to this low-level long-standing computer vision task.
Dong \etal~\cite{dong2015compression} firstly proposed a four-layer network named ARCNN to reduce JPEG compression artifacts. Afterwards, \cite{li2017efficient,mao2016image,svoboda2016compression,tai2017memnet,zhang2017beyond, zhang2017learning} proposed deeper networks to further reduce compression artifacts. 
A notable example is \cite{zhang2017beyond}, which devised an end-to-end trainable denoising convolutional neural network (DnCNN) for Gaussian denoising
. DnCNN also achieves a promising result on JPEG deblocking task. Moreover, \cite{chen2018dpw,guo2016building,yoo2018image} enhanced visual quality by exploiting the wavelet/frequency domain information of JPEG compression images. Recently, more methods \cite{bigdeli2017deep,chang2017one,cui2018decoder,galteri2017deep,guo2017one,liu2018non,maleki2018blockcnn} were proposed and got competitive results. Concretely, Galteri \etal~\cite{galteri2017deep} used a deep generative adversarial network and recovered more photorealistic details. 
Inspired by \cite{wang2018non}, \cite{liu2018non} incorporated non-local operations into a recursive framework for quality restoration, it computed self-similarity between each pixel and its neighbors, and applied the non-local module recursively for correlation propagation. 
Differently, here we adopt a non-local module to capture global motion patterns by exploiting inter-frame pixel-wise similarity.

\subsection{Video Compression Artifact Reduction}
Most of existing video compression artifact reduction works~\cite{dai2017convolutional,jin2018quality,wang2017novel,yang2018enhancing,yang2017decoder} focus on individual frames, neglecting  the spatiotemporal correlation between neighboring frames.
Recently, several works were proposed to exploit the spatiotemporal information from neighboring frames. Xue \etal~\cite{xue2019video} designed a neural network with a motion estimation component and a video processing component, and utilized a joint training strategy to handle various low-level vision tasks
. Lu \etal~\cite{lu2018deep} further incorporated quantized prediction residual in compressed code streams  as strong prior knowledge, and proposed a deep Kalman filter network (DKFN) to utilize the spatiotemporal information from the preceding frames of the target frame.
In addition, considering that quality of nearby compressed frames fluctuates dramatically, 
\cite{guan2019mfqe,yang2018multi} proposed multi-frame quality enhancement (MFQE) and utilized motion compensation of two nearest PQFs 
to enhance low-quality frames. Comparing with DKFN\cite{lu2018deep}, MFQE is a post-processing method and uses less prior knowledge of the compression codec, but still achieves state-of-the-art performance on HEVC compressed videos.

In addition to compression artifact removal, spatiotemporal correlation mining is also a hot topic in other video quality enhancement tasks, such as video super resolution~(VSR). \cite{caballero2017real,kappeler2016video,kim2018spatio, liu2017robust,sajjadi2018frame,tao2017detail,xue2019video} estimated optical flow and warped several frames to capture the hidden spatiotemporal dependency for VSR. 
Although these methods work well, they rely heavily on the accuracy of motion estimation. Instead of explicitly taking advantage of motion between frames, \cite{jo2018deep} utilized a 3D convolutional network as a dynamic filter generation network to generate dynamic upsampling filter and fine residual for VSR.

In summary, most end-to-end CNN based visual quality enhancement methods consider only either a single frame or a pair of neighboring frames, thus may miss important details of other neighboring frames. 
Unlike these works, here we employ the NL-ConvLSTM mechanism to utilize multiple frames and capture spatiotemporal variations in a frame sequence without explicit motion estimation and compensation, 
With an approximate strategy for non-local similarity computation, our method can effectively reduce artifacts and achieves state-of-the-art performance.

\section{Method}
The goal of video compression artifact reduction is to infer a high quality frame ${{\hat{Y}}_t}$ from a compressed frame ${X}_t$ of the original frame~(ground truth) ${Y}_t$, where ${X}_t $$\in$$ \mathbb{R}^{C\times N}$ is the compressed frame at time $t$. Here, $C$ is the number of channels of a single frame. For the sake of notation clarity, we collapse the spatial positions~(width $W$ and height $H$) into one dimension, $N$$=$$HW$.  
Let ${\mathcal{X}}_t$$=$$\{{X}_{t-T},\dots,{X}_{t+T}\}$ denote a sequence of  ($2T+1$) consecutive compressed frames, our method takes ${\mathcal{X}}_t$ as input and outputs ${\hat{Y}}_t$.

\subsection{The Framework}

Our method is an end-to-end trainable framework that consists of three modules: Encoder, NL-ConvLSTM module and Decoder, as shown in Fig.~\ref{fig:architecture}. They are respectively responsible for extracting features from individual frames, learning spatiotemporal correlation across frames, 
and decoding high-level features to residuals, with which the high-quality frames are reconstructed eventually. 

\textbf{Encoder}. It 
is designed with several 2D convolutional layers to extract features from ${\mathcal{X}}_t$. With ${\mathcal{X}}_t$ as input, it outputs ${\mathcal{F}_t}$$=$$\{{F}_{t-T},\dots,{F}_{t+T}\}$. 
Here, $F_t $$\in$$ \mathbb{R}^{C_f \times N}$ is the corresponding feature extracted from $X_t$, $C_f$ is the channel size of output features. It processes
each frame individually.

\textbf{NL-ConvLSTM}. To trace the spatiotemporal dependency in a frame sequence, 
we put a ConvLSTM~\cite{xingjian2015convolutional} module between the Encoder and the Decoder.
ConvLSTM is able to capture spatiotemporal information from a frame sequence of arbitrary length, but it is not good at handling large motions and blurring motions well~\cite{kappeler2016video}. 
To tackle this problem, 
we embed the non-local (NL)~\cite{buades2005non} mechanism into ConvLSTM, and develop the NL-ConvLSTM module. Here, the non-local similarity is used for pixels from different frames rather than pixels within a frame~\cite{buades2005non}. The NL-ConvLSTM module $\mathcal{N}$ can be described as
\begin{equation}
\left[ \mathcal{H}_t, \mathcal{C}_t \right] = \mathcal{N} \left( F_{t-1}, F_{t}, [\mathcal{H}_{t-1}, \mathcal{C}_{t-1}] \right ).
\end{equation}
Different from ConvLSTM in \cite{tao2017detail,xingjian2015convolutional} that is fed with only feature ${F}_t$ at time $t$, NL-ConvLSTM takes additional feature ${F}_{t-1}$ at time ($t$-$1$) as input, and outputs the corresponding hidden state and cell state $\mathcal{H}_t, \mathcal{C}_t \in \mathbb{R}^{C_h \times N}$. Here, $C_h$ is the number of channels of hidden state and cell state. 
Moreover, hidden state $\mathcal{H}_{t-1}$ and cell state $\mathcal{C}_{t-1}$ are not fed into gate operation directly in NL-ConvLSTM.
In contrast, we calculate inter-frame pixel-wise similarity $S_t$ between $F_{t-1}$ and $F_t$, then perform a weighted sum over $\mathcal{H}_{t-1}$ and $\mathcal{C}_{t-1}$ with $S_t$ as weight. In addition, bi-directional ConvLSTM is used in our paper to learn spatiotemporal dependency from both preceding and following frames. In the following sections, we only mention the operation of forward NL-ConvLSTM for simplicity. Details of the NL-ConvLSTM module can be referred to Fig.~\ref{fig:architecture}~(right), Fig.~\ref{fig:difference}, Section \ref{sec:non-local} and \ref{sec:two-stage}.

\textbf{Decoder}. It decodes the hidden state from both directions of NL-ConvLSTM module into residual, with which the high-quality frame is reconstructed. Specifically, we first combine the hidden states by a convolutional layer whose kernel size is 1$\times$1, then use several stacked convolutional layers to generate the residual.

\subsection{Non-local ConvLSTM}
\label{sec:non-local}

ConvLSTM can be described as follows
\cite{tao2017detail}: 
\begin{equation}
[\mathcal{H}_t, \mathcal{C}_t] = ConvLSTM(F_t, [\mathcal{H}_{t-1}, \mathcal{C}_{t-1}]).
\end{equation}
To learn robust spatiotemporal dependency
, we adopt the non-local mechanism into ConvLSTM to help estimate motion patterns in frame sequences.
As an extension of ConvLSTM, NL-ConvLSTM can be formulated as
\begin{equation}
\begin{split}
S_t &= NL\left (F_{t-1},F_t \right ), \\
\left [\hat{\mathcal{H}}_{t-1}, \hat{\mathcal{C}}_{t-1} \right ]&= NLWarp(\left [\mathcal{H}_{t-1}, \mathcal{C}_{t-1} \right ], S_t),\\
\left [H_t, C_t \right ] &= ConvLSTM(F_{t}, \left [\hat{\mathcal{H}}_{t-1}, \hat{\mathcal{C}}_{t-1} \right ]), \\
\end{split}
\end{equation}
where $S_t \in \mathbb{R}^{N\times N}$ denotes the similarity matrix between the pixels of the current frame and all pixels of the preceding frame. 
$NL$ is the non-local operator for calculating the similarity matrix between features of two frames, $NLWarp$ is the warping operation for the hidden state and cell state at time ($t$-1) with a weighted sum form.

Following the non-local operation~\cite{buades2005non}, the inter-frame pixel-wise similarity and non-local warping operation in our work are as follows:
\begin{equation}
\label{equ:nonlocal}
\begin{split}
D_t \left ( i, j \right ) &= \left \| F_{t-1} \left ( i \right )  - F_t \left (  j \right ) \right \|_2, \\
S_t \left ( i, j \right ) &= \frac{\exp{ \left ( -
		D_t \left ( i, j \right )/\beta
		\right )} }{\sum\nolimits_{\forall i}{\exp{ \left (-
			D_t \left ( i, j \right )/\beta
			\right )}}},  \\
\left [\hat{\mathcal{H}}_{t-1}, \hat{\mathcal{C}}_{t-1} \right ] &=
\left[\mathcal{H}_t\cdot S_t, \mathcal{C}_t \cdot S_t \right], \\
\end{split}
\end{equation}
where $i,j\in\left\{1,\cdots,N\right\}$ are indices of pixels in a feature map, $F(i)$ and $\mathcal{H}(i)$ are the corresponding feature and state at position $i$. $D_t\left ( i, j \right )$ and $S_t\left ( i, j \right )$ are the Euclidean distance and similarity throughout all channels between pixel $i$ in the preceding feature map at time $t$-$1$ and pixel $j$ in the current feature map at time $t$. $S_t\left ( i, j \right )$ satisfies $\sum\nolimits_i{S_t \left ( i, j \right ) = 1}$. 
Thus, the non-local method can be seen as a special attention mechanism~\cite{wang2018non}.

\subsection{Two-stage Non-local Similarity Approximation}
\begin{figure}[t]
	\begin{center}
		\includegraphics[width=\linewidth]{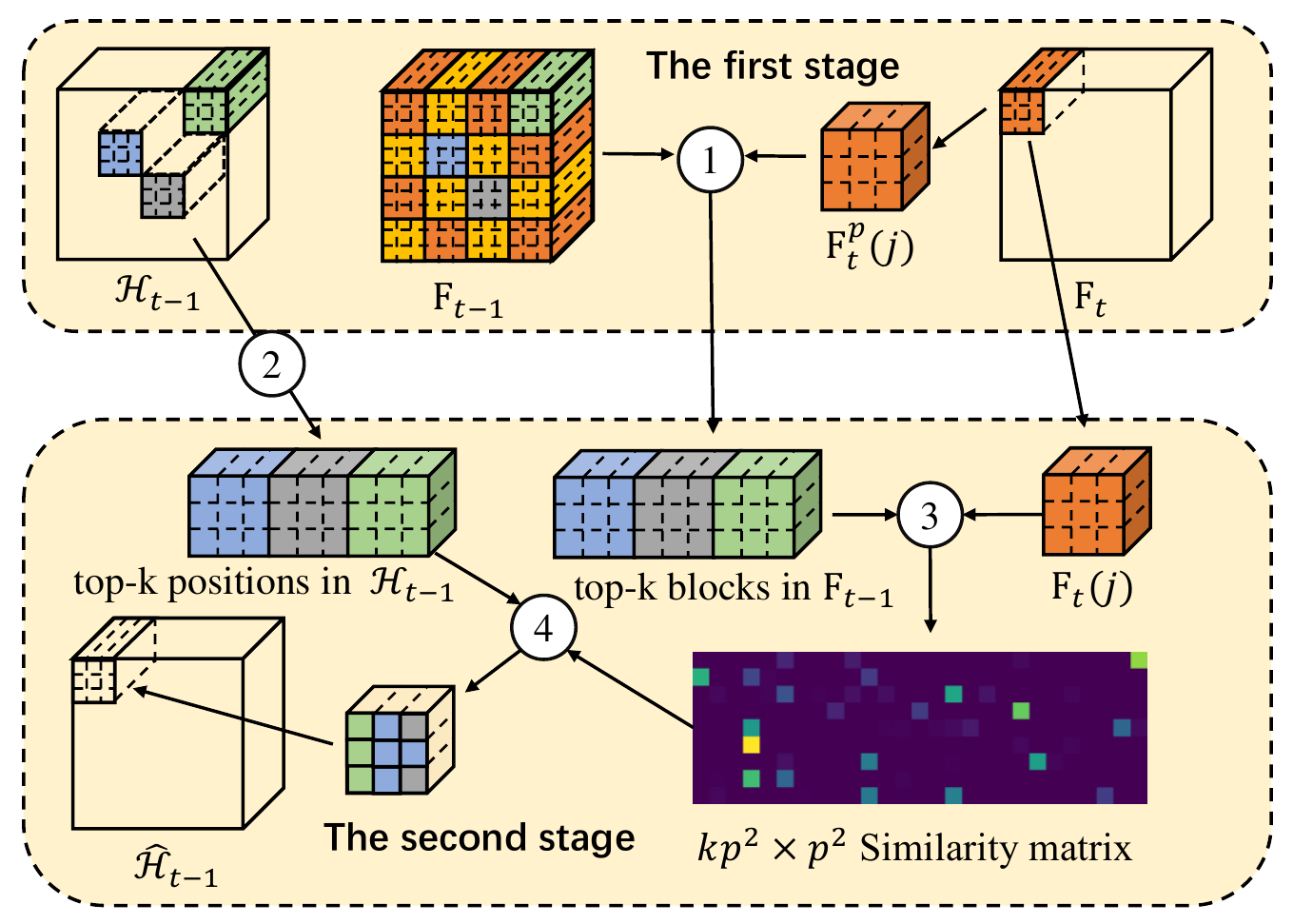}
	\end{center}
	\vspace{-5pt}
	\caption{The workflow of two-stage similarity approximation. \ding{172} finding the top-$k$ most similar blocks in $F_{t-1}$ with respect to block $F_t^p(j)$ from $F_t$; \ding{173} extracting blocks in $\mathcal{H}_{t-1}$ from the corresponding positions of the top-$k$ most similar blocks in $F_{t-1}$; \ding{174} calculating pixel-wise similarity between the selected blocks from $F_{t-1}$ and $F_t^p(j)$; \ding{175} NLWarp operation for $\mathcal{H}_t$. }
	\label{fig:difference}
	
\end{figure}

\label{sec:two-stage}
Directly computing $S_t$$ \in $$\mathbb{R}^{N \times N}$ and the warping operation will incur extremely high computation and memory cost for high-resolution videos.
Therefore, we propose a two-stage non-local method to approximate $D_t$ as $\hat{D}_t$ and $S_t$ as $\hat{S}_t$, which can reduce computation and memory while maintaining accuracy. 
The key idea of our approximate method is to pre-filter image blocks according to the deep feature learned by the Encoder before calculating pixel-wise similarity. 
The details are as follows:

In the $1^{st}$ stage, we use average pooling to downsample 
the feature map from the Encoder and reduce the block matching sensitivities of geometry transformations (shifts and rotations). 
Denote the kernel size of the average pooling as $p$ and the downsampled feature map as $F_t^p$. Then, the resolution of the feature map is reduced to 
$N/p^2$, i.e., $1 / p^2$ of the original resolution. Each super-pixel in the downsampled 
feature map $F_t^p$ corresponds to a block consisting of $p^2$ pixels in the original feature map. 
Thus, the downsampled distance matrix $D_t^p \in \mathbb{R}^{(N/p^2)^2 }$ can be calculated by
\begin{equation}
D_t^p \left(i,j \right) = \left \| F_{t-1}^p \left ( i \right )  - F_t^p \left (  j \right ) \right \|_2.
\end{equation}
For each pixel in any block $b_t$ of $F_t^p$, we consider only $k$$\times$$p^2$ pixels in the $k$ blocks of $F_{t-1}^p$ that are nearest to $b_t$. 

In the $2^{nd}$ stage, we compute and store the 
similarities between each pixel of $F_{t}$ 
and the corresponding $k$$\times$$p^2$ pixels of $F_{t-1}$. 
While for the other pixels in the preceding frame, the elements of $\hat{D}_t$ and $\hat{S}_t$ are set to $+\infty $ and $0$ respectively. As similar pixels are sparse and a pixel can be represented by a few pixels from the neighboring frame, the quality loss of top-$k$ blocks approximation is negligible. 
Fig.~\ref{fig:difference} illustrates the workflow of two-stage similarity approximation. The NLWarp operation for $\mathcal{C}_t$ is similar to that of $\mathcal{H}_t$. We show  only the operation for $\mathcal{H}_t$ in Fig.~\ref{fig:difference} for simplicity.

\textbf{Complexity analysis.} Tab.~\ref{table:computation complexity} compares the complexity of our approximate method with that of the original method. Since $\log k \ll C$ in our experiments, we can neglect the item related to  searching the top-$k$ nearest blocks in Tab.~\ref{table:computation complexity} for simplicity. We denote the complexity of the original non-local method as $\psi$, and the complexity of our approximate method as $\phi$, which can be rewritten to $\mathcal{O}((N/p^2)^2C+2kNCp^2)$.
By properly choosing the values of $k$ and $p$ so that $kp^2 \ll N$, we have $\phi / \psi = 1/(2p^4)+kp^2/N \ll 1$, which means that our method dramatically reduces the computation cost of the original method. And for a given $k$, $\phi / \psi$ achieves the minimum $1.5(k/N)^{2/3}$ with $p$=$(N/k)^{1/6}$. Similar conclusion can be drawn for memory cost. 
More specifically, by setting $p$=$10$, $k$=$4$, $C$=$64$, and $f$=$4$\footnote{$f$ is the kernel size of a convolutional layer}, $\phi$ is close to ($\mathcal{O}(NC^2f^2)$), which is the computational complexity of a convolutional layer with a $f$$\times $$f$ kernel.

\textbf{Non-local operation \emph{vs.} motion compensation. } There are similarities and differences between the two operations. \textit{Similarities}: 1) Both can be adopted to capture spatiotemporal relationships and motion patterns in consecutive frames. 2) Both can be seen as an attention mechanism. In non-local operation, the warped state $\hat{\mathcal{H}}_{t-1}$ is calculated with the states of all pixels in $\mathcal{H}_{t-1}$ in a weighted sum form; 
while in motion compensation, each pixel in $\hat{\mathcal{H}}_{t-1}$ is evaluated by interpolation with some neighboring pixels in $\mathcal{H}_{t-1}$, 
also in a weighted sum form. 
\textit{Differences}: 1) In non-local operation, each pixel is warped from multiple positions in $\mathcal{H}_{t-1}$, and the motion is not restricted by a fixed flow magnitude, which is different from motion compensation where a fixed flow magnitude has to be set. Thus, non-local operation can capture global motion patterns more effectively.
2) In non-local operation, the similarity is determined once the feature is extracted; while in motion compensation, we need to train additional layers for motion field generation.

\subsection{Implementation Details}

In our implementation, following existing methods~\cite{lu2018deep,yang2018multi}, we use $L_2$ norm as the loss function:
\begin{equation}
l\left({\mathcal{X}}_t, Y_t\right) = \left \| \hat{Y}_t  - Y_t \right \|_2.
\end{equation}

Due to the advantage of NL-ConvLSTM, global motion can be captured with a small kernel in ConvLSTM. So our NL-ConvLSTMs is implemented with a $3 \times 3$ kernel.
For all datasets, the networks are trained using the ADAM~\cite{kingma2014adam} optimizer with an initial learning rate of $10^{-4}$ and a mini-batch size of 32. In training, raw and compressed sequences are sampled with a patch size of $80 \times 80$ for NL-ConvLSTM. In contrast, the full resolution video sequences are fed into our model during testing.

We use $k$=$4$ and $p$=$10$ in all experiments for balancing efficiency and effectiveness, and set $T$=$3$ for all datasets.

To further accelerate the non-local operator, vectorization is adopted in calculating distance matrix $D_t$. 
Although vectorization does not reduce the number of floating-point operations, it enables acceleration via parallel computing.
By expanding $D_t$ in Equ.~(\ref{equ:nonlocal}), we have
\begin{equation}
\begin{split}
{D_t}^2 = {\sum^{C_f}{{F}_{t-1}^2} \cdot \bold{1}^\top +
	\bold{1} \cdot \sum^{C_f}{{F_t^2}}^\top
	- 2F_{t-1}^\top \cdot F_t}, \\
\end{split}
\label{equ:vectorize}
\end{equation}
where 
$\bold{1} \in \mathbb{R}^{N \times 1}$ is a vector whose elements are $1$. We adopt Equ.~(\ref{equ:vectorize}) for calculating $D_t^p$ in the $1^{st}$ stage, and implement a sparse version of Equ.~(\ref{equ:vectorize}) to compute distances between each pixel in the current frame and $k \times p^2$ pre-filtered pixels in the preceding frame.

\begin{table}
	%
	%
	\caption{{
			Complexity comparison of the original non-local approach and ours. Here, $N$ and $C$ are the numbers of positions and channels, $k$ and $p$ are the number of pre-filtered blocks and the downsampling scale. By setting $k$=$4$ and $p$=$10$, our method cuts the time and space to about 1/1000 of that consumed by the original non-local method in 1080P videos. }}
	\centering
	\scriptsize
	\begin{tabular}[h]{lcc}
		\toprule
		& Original non-local  & NL-ConvLSTM \\
		\cmidrule(r){2-3}
		Time        & $\mathcal{O}(2N^2C)$                          & $\mathcal{O}((N/p^2)^2(C+\log k)+2kNCp^2)$   \\
		Space       & $\mathcal{O}(2N^2)$                          & $\mathcal{O}((N/p^2)^2+kN/p^2+2kNp^2)$   \\
		\bottomrule
		\label{table:computation complexity}
	\end{tabular}
\end{table}

\section{Performance Evaluation}
To evaluate our method, we conduct extensive experiments on two datasets: Vimeo-90K~\cite{xue2019video} and Yang \etal's dataset~\cite{yang2018multi}. Our evaluation consists of five parts: 1)~Ablation study; 2)~Quantitative evaluation with two performance metrics~(PSNR and SSIM)
; 3)~Qualitative evaluation by comparing the visual effect of compression artifact reduction of different methods; 4)~Run time comparison; 5)~Checking the effectiveness of our method on videos compressed by another algorithm.

\subsection{Datasets and Settings}
\textbf{Vimeo-90K.} It is a recently-built large-scale video dataset for low-level video processing
. All frames are resized to a fixed resolution $448 \times 256$. We follow the settings in \cite{lu2018deep} and refer the interested readers to \cite{lu2018deep} for details. 
In short, compressed clips are generated by x265 in FFmpeg with quantization parameter $QP$=32 and 37. Loop filter, SAO~\cite{sullivan2012overview} and B-frames are disabled in codec.
We follow \cite{lu2018deep} and only evaluate the $4^{th}$ frame of each clip.

\begin{figure*}
	\begin{center}
		\includegraphics[width=\linewidth]{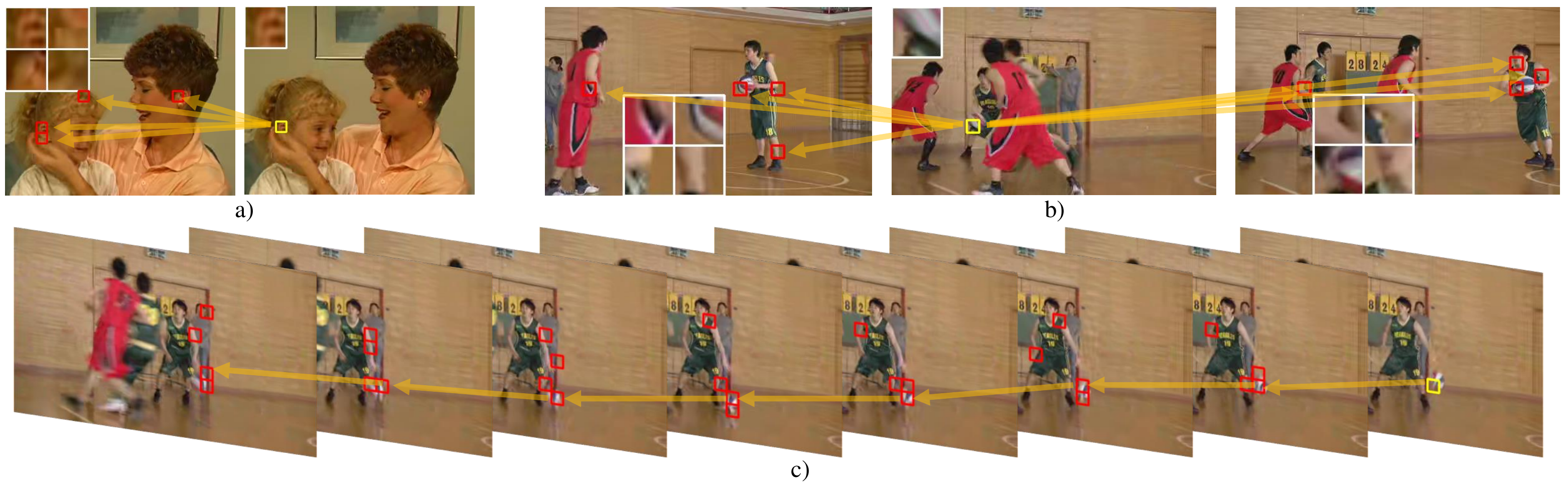}
	\end{center}
	\vspace{-5pt}
	\caption{Examples of blocks after pre-filtering in the first stage of our method. Images are from Yang \etal's dataset. The red blocks are the top-$4$ most similar blocks with respect to the yellow block in another frame.}
	\label{fig:nonlocal example}
\end{figure*}

\textbf{Yang \etal's dataset.}  It consists of 70 video sequences selected from the datasets of Xiph.org\footnote{https://media.xiph.org/video/derf/} and JCT-VC~\cite{bossen2012common}. Resolutions of these video sequences vary from $352 \times 240$ to $2,560 \times 1,600$. For a fair comparison, we follow the settings in \cite{yang2018multi}: $60$ sequences are taken for training and the remaining $10$ for testing. 
All sequences are encoded in HEVC LDP mode, using HM 16.0 with $QP$=37 and 42.

\begin{table}
	%
	%
	\caption{{
			Ablation study of the proposed NL-ConvLSTM on Yang \etal's dataset with $QP$=37. The results of PSNR improvement $\Delta$PSNR~(db) are reported in the $1^{st}$ row. The results of SSIM improvement $\Delta$SSIM~($\times$$10^{-2}$) are listed in the $2^{nd}$ row.
	}}
	\centering
	\scriptsize
	\begin{tabular}[h]{l@{\hspace{0.3em}}c@{\hspace{0.5em}}c@{\hspace{0.5em}}c@{\hspace{0.5em}}c}
		\toprule
		& Encoder-Decoder  & ConvLSTM & ME-ConvLSTM & Our method \\
		& with 1 frame & with 7 frames & with 7 frames & with 7 frames \\
		\cmidrule(r){2-5}
		$\Delta$PSNR       &  0.395   & 0.456 &  0.503  &  \textbf{0.601} \\
		$\Delta$SSIM       &  0.684  & 0.723  &  0.827  &  \textbf{0.897} \\
		\bottomrule
		\label{tab:ablation}
	\end{tabular}
\end{table}

\subsection{Ablation Study}
\textbf{Effect of multiple frames.} Here we evaluate the effectiveness of utilizing multiple frames. First, we compare the performance between Encoder-Decoder (using 1 frame) and  ConvLSTM (using 7 frames). The results of the 2nd and 3rd columns in Table~\ref{tab:ablation} show that utilizing multiple frames in ConvLSTM obviously boosts performance. Then, we further verify the effectiveness of our method using more frames as input on Yang \etal's dataset with $QP$=37. Each of the input clips consists of 20 consecutive frames: the target frame, 15 preceding frames and 4 following frames. We use the pre-trained model on the same settings with $T$=3, and then fine-tune the model on 20-frame inputs using almost similar training settings, except for a smaller batch size. We find that our model tuned on longer sequences gets 0.604dB/0.00923 of $\Delta$PSNR/$\Delta$SSIM, which are better than the results of the 5th column in Table~\ref{tab:ablation}
. The results above indicate that multiple frames and longer sequences do improve artifact reduction performance.

\textbf{Effect of the non-local mechanism.} Here we investigate the effectiveness of non-local mechanism. The non-local module aims to learn  spatiotemporal dependency between two consecutive frames. Usually, motion estimation and compensation module can do such a role. Thus, we use a motion estimation and compensation module to replace the non-local module, and name such a method as ME-ConvLSTM. We follow the motion generation architecture in \cite{shi2017deep}. The results on Yang \etal's dataset with $QP$=37 are listed in the 4th column of Table~\ref{tab:ablation}. Compared with ME-ConvLSTM, our method performs better, $19.48\%$ higher in terms of $\Delta$PSNR and $8.46\%$ higher in terms of $\Delta$SSIM, which demonstrates that ConvLSTM with non-local mechanism utilizes temporal information better.


\textbf{Effect of the Two-Stage Non-local Method.} We propose a two-stage non-local method to learn the spatiotemporal dependency between two neighboring frames. In the first stage, it tries to find the top-$k$ most similar blocks in $F_{t-1}^{p}$ for each block of $F_{t}^p$. This pre-filtering impacts spatiotemporal dependency learning. Here, we visualize some pre-filtering results in the first stage to show that our NL-ConvLSTM method can learn the spatiotemporal dependency among consecutive frames.

Fig.~\ref{fig:nonlocal example} shows some examples of blocks after pre-filtering. The red blocks are the top-$4$ most similar blocks in a frame after pre-filtering with respect to the yellow block in another frame. In Fig.~\ref{fig:nonlocal example}(a) and Fig.~\ref{fig:nonlocal example}(b), these blocks are enlarged, bounded with white boxes, and shown in the full image. Fig.~\ref{fig:nonlocal example}(a) shows two consecutive frames. The yellow block in the right frame contains the daughter's ear. Our method finds the daughter's ear and her mother's ear in the left frame. 
This shows that our method can capture similar patterns at various locations, which could provide additional information for enhancement. In Fig.~\ref{fig:nonlocal example}(b), we manually construct a 3-frame sequence with a large time gap. It is difficult for motion estimation based methods to handle such large motion. However, our method can still capture the spatiotemporal dependency of basketball.
In Fig.~\ref{fig:nonlocal example}(c),  we thread the top-1 block of each frame from the yellow block in the right-most frame to left iteratively. Such a path reflects the robust spatiotemporal relationship built among frames by our method.

\begin{table}
	\centering
	\caption{Average PSNR/SSIM on Vimeo-90K.}
	\label{tab:vimeo}
	\renewcommand\tabcolsep{4.5pt}
	\begin{tabular}{c|cc}
		\toprule
		\tabincell{c}{QP} &
		\tabincell{c}{32} &
		\tabincell{c}{37} \cr
		\midrule
		HEVC ~\cite{sullivan2012overview} & 34.19 / 0.950 & 31.98 / 0.923 \cr
		ARCNN ~\cite{dong2015compression} & 34.87 / 0.954 & 32.54 / 0.930 \cr
		DnCNN ~\cite{zhang2017beyond} & 35.58 / 0.961 & 33.01 / 0.936 \cr
		DSCNN ~\cite{yang2017decoder} & 35.61 / 0.960  & 32.99 / 0.938 \cr
		DKFN ~\cite{lu2018deep} & 35.81 / 0.962 & 33.23 / 0.939 \cr
		3D CNN &  35.81 / 0.961 & 33.25 / 0.938 \cr
		Our method & \bf{35.95} / \bf{0.965} & \bf{33.39} / \bf{0.943} \cr
		
		\bottomrule
	\end{tabular}
\end{table}

\subsection{Quantitative Comparison}

For a fair comparison, we use the same data processing method and training data. 
Thus, to demonstrate the advantage of our method, we 
compare it with five existing methods: ARCNN~\cite{dong2015compression}, DnCNN~\cite{zhang2017beyond}, DSCNN~\cite{yang2017decoder}, DKFN~\cite{lu2018deep} and MFQE~\cite{yang2018multi}.
For ARCNN, DnCNN and DSCNN, we get better results after they were retrained on Yang \etal's dataset than that reported in \cite{yang2018multi}. 
For DKFN~\cite{lu2018deep}, we directly cite performance results from the original paper where it was evaluated only on Vimeo-90K. 
For MFQE~\cite{yang2018multi}, we cite the results of PSNR improvement, and compute SSIM improvement via our manually labeled PQFs and its published model\footnote{\cite{yang2018multi} only publishes the model with $QP$=37.}.
Besides, considering that 3D CNN is able to capture spatiotemporal information from video frames, we also implement it for performance comparison by adopting the same architecture as \cite{jo2018deep} that aims at video super-resolution. Due to its large memory cost, we train and evaluate it only on Vimeo-90K.

\begin{table}
	\centering
	\scriptsize
	\fontsize{6.5}{9}\selectfont
	\caption{Average $\Delta$PSNR~(dB) and $\Delta$SSIM~($\times10^{-2}$) on Yang \etal's dataset.}
	\begin{tabular}{|@{\hspace{0.4em}}c@{\hspace{0.4em}}|c|@{\hspace{0.25em}}c@{\hspace{0.25em}}|@{\hspace{0.25em}}c@{\hspace{0.25em}}|@{\hspace{0.25em}}c@{\hspace{0.25em}}|@{\hspace{0.25em}}c@{\hspace{0.25em}}|@{\hspace{0.25em}}c@{\hspace{0.25em}}|}
		\hline
		
		QP  & Seq.  & ARCNN~\cite{dong2015compression} & DnCNN~\cite{zhang2017beyond} & DSCNN~\cite{yang2017decoder} & MFQE~\cite{yang2018multi}  & Our method \\
		\hline
		\multirow{11}[0]{*}{37} & 1     &  0.241 / 0.51     &  0.448 / 0.83     &  0.492 / 0.87     &   0.772 / 1.15    &  \textbf{0.827} / \textbf{1.21}  \\
		\cline{2-7}          &         2     &   0.115 / 0.30    &   0.439 / 0.52    &   0.458 / 0.58    &    0.604 / 0.63  &     \textbf{0.971} / \textbf{0.92}     \\
		\cline{2-7}          &         3     &   0.161 / 0.49    &   0.276 / 0.76    &   0.271 / 0.74    &    0.472 / 0.91  &     \textbf{0.483} / \textbf{0.99}      \\
		\cline{2-7}          &         4     &   0.183 / 0.35    &   0.377 / 0.55    &   0.393 / 0.54    &    0.438 / 0.48  &     \textbf{0.576} / \textbf{0.66}      \\
		\cline{2-7}          &         5     &   0.150 / 0.30    &   0.333 / 0.48    &   0.356 / 0.53    &    0.550 / 0.52  &     \textbf{0.598} / \textbf{0.74}      \\
		\cline{2-7}          &         6     &   0.161 / 0.23    &   0.415 / 0.50    &   0.435 / 0.49    &    0.598 / 0.51  &     \textbf{0.658} / \textbf{0.67}      \\
		\cline{2-7}          &         7     &   0.128 / 0.29    &   0.284 / 0.44    &   0.277 / 0.45   &    0.390  / 0.45 &     \textbf{0.394} / \textbf{0.58}      \\
		\cline{2-7}          &         8     &   0.125 / 0.37    &   0.276 / 0.61    &   0.230 / 0.63    &    0.484 / 1.01   &     \textbf{0.563} / \textbf{1.18}      \\
		\cline{2-7}          &         9     &   0.149 / 0.38    &   0.299 / 0.71    &   0.271 / 0.66    &    0.394 / 0.92  &     \textbf{0.439} / \textbf{1.03}      \\
		\cline{2-7}          &        10     &   0.146 / 0.24    &   0.289 / 0.58    &   0.274 / 0.54    &    0.402 / 0.80  &     \textbf{0.501} / \textbf{0.99}      \\
		\cline{2-7}          &  \bf{Ave.}    &   0.156 / 0.35    &   0.344 / 0.59    &   0.346 / 0.60    &    0.510 / 0.74  &     \textbf{0.601} / \textbf{0.90}     \\
		\hline
		42               & \bf{Ave.}     &   0.252 / 0.83    &   0.301 / 0.96    &    0.364 / 1.06   &    0.461 / ---   &   \textbf{0.614} / \textbf{1.47}       \\
		\hline
		\multicolumn{7}{c}{1: \textit{PeopleOnStreet}\ \ 2: \textit{TunnelFlag}\ \ 3: \textit{Kimono}\ \ 4: \textit{BarScene}\ \ 5: \textit{Vidyo1}}\\
		\multicolumn{7}{c}{6: \textit{Vidyo3}\ \ 7: \textit{Vidyo4}\ \ 8: \textit{BasketballPass}\ \ 9: \textit{RaceHorses}\ \ 10: \textit{MaD}}
	\end{tabular}%
	\label{tab:yang}%
\end{table}%

\textbf{Quality enhancement.}
Results of PSNR/SSIM on two datasets are in Table~\ref{tab:vimeo} and Table~\ref{tab:yang} respectively.
In Table~\ref{tab:vimeo}, we use HEVC with loop filters  ~\cite{sullivan2012overview} as the baseline.
From Table~\ref{tab:vimeo}, we can see that our method outperforms the $2^{nd}$ best methods (DKFN and 3D CNN) by about 0.14 dB in terms of PSNR and $25\%$ improvement on $\Delta$SSIM. These methods that utilize spatiotemporal information, including ours, DKFN and 3D CNN, all  achieve better performance than the remaining methods.

Yang \etal's dataset has less training data than Vimeo-90K, and these two datasets are processed with different compression settings. All methods perform a little worse on Yang \etal's dataset than on Vimeo-90K. However, from Table~\ref{tab:yang} we can still get similar conclusion: methods exploiting spatiotemporal information of neighboring frames perform better than these do not exploit. Our method outperforms all the other methods on all test sequences, and on average its PSNR/SSIM improvement is $17.8\%$/$21.6\%$ higher than that of MFQE. Specifically, for $QP$=37, our method achieves the highest PSNR/SSIM improvement on the $2^{nd}$ sequence, which is $60.7\%$ / $46\%$ higher than that of MFQE. 
For $QP$=42, our method gets $33.19\%$ and $68.68\%$ increase of $\Delta$PSNR over MFQE and DSCNN, respectively.
\begin{figure}
	\begin{center}
		\includegraphics[width=\linewidth]{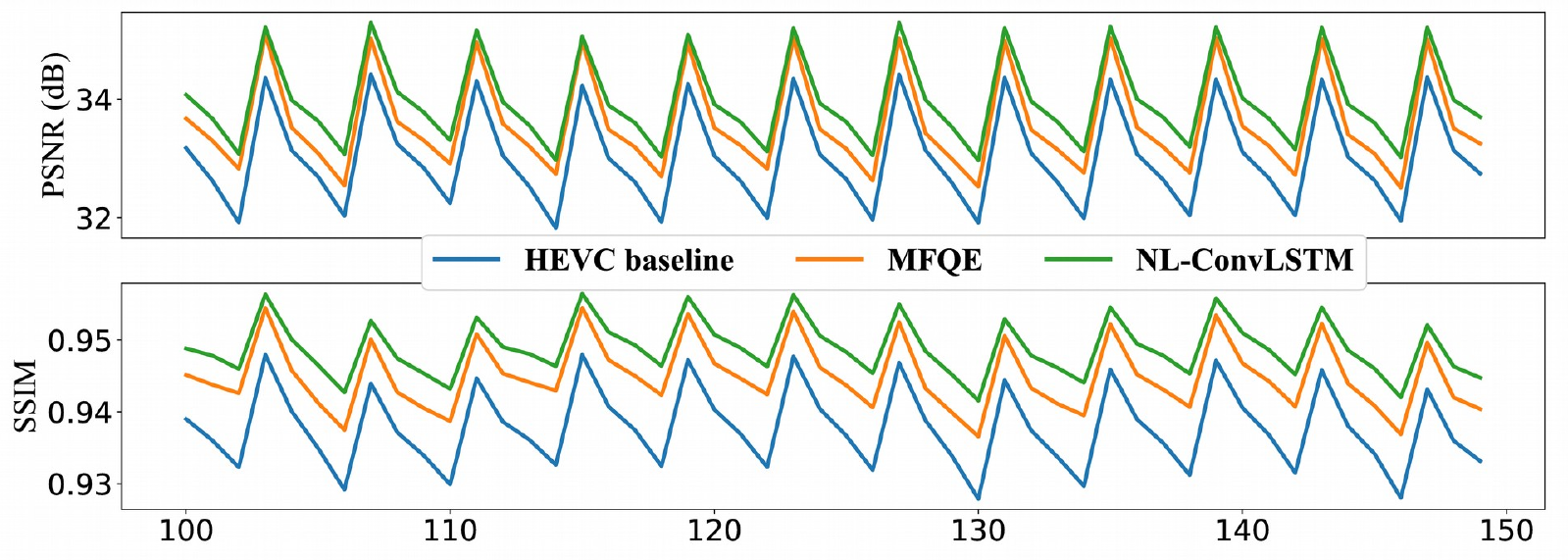}
	\end{center}
	\vspace{-6pt}
	\caption{PSNR/SSIM curves of HEVC baseline, MFQE and NL-ConvLSTM on the sequence \emph{TunnelFlag} with $QP$=37.}
	\label{fig:std}
\end{figure}

\begin{figure*}
	\begin{center}
		\includegraphics[width=\linewidth]{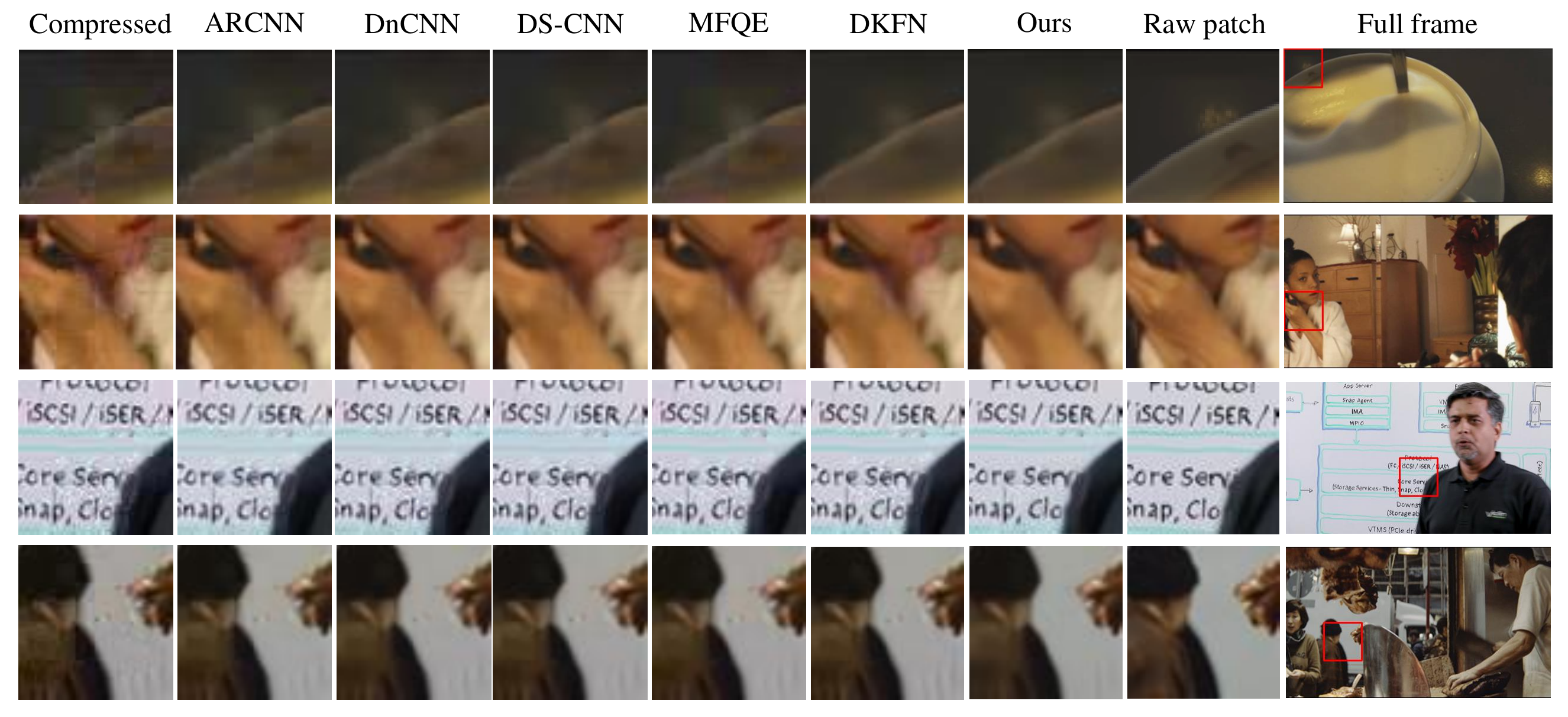}
	\end{center}
	\vspace{-5pt}
	\caption{Comparison of reduction effect on four kinds of compression artifacts.}
	\label{fig:artifact}
\end{figure*}

\begin{table}
	\caption{Run-time (\emph{ms per frame}) comparison among six methods.}
	\scriptsize
	\begin{tabular}{|@{\hspace{0.25em}}c@{\hspace{0.25em}}|c|c|c|c|c|}
		\hline
		Resolution & 180x180 & 416x240 & 640x360  & 1280x720 & 1920x1080 \cr
		\hline
		ARCNN~\cite{dong2015compression} & 1.73 & 4.58 & 9.19 & 36.06 & 80.70 \cr
		\hline
		DnCNN~\cite{zhang2017beyond} & 6.30 & 15.84 & 35.51 & 139.77 & 315.83 \cr
		\hline
		DSCNN~\cite{yang2017decoder} & 15.26 & 36.88 & 82.31 & 322.92 & 731.21 \cr
		\hline
		MFQE\tablefootnote{time for detecting PQFs are not included}~\cite{yang2018multi} & 20.28+ & 51.01+ & 112.87+ & 443.82+ & 1009.00+\cr
		\hline
		original NL & 4391.75 & - & - & - & - \cr
		\hline
		ours & 102.13 & 304.11 & 621.94 & 2607.60 & 6738.00\\
		\hline
	\end{tabular}
	\label{run time}
	\vspace{-3pt}
\end{table}

\textbf{Quality fluctuation.} Quality fluctuation is an index to evaluate the quality of a whole video~\cite{guan2019mfqe,yang2018multi}. Drastic quality fluctuation often leads to severe temporal inconsistency and degradation of QoE. We evaluate fluctuation by Standard Deviation~(STD) and Peak-Valley Difference~(PVD)\footnote{PVD calculates the average difference between peak values and their preceding/following nearest valley values} of the PSNR/SSIM curves for each sequence as in \cite{guan2019mfqe,yang2018multi}. 
Here, we present only the STDs and PVDs of HEVC baseline, MFQE and our method for simplicity. For PSNR, the STD values of HEVC baseline, MFQE and our method are $1.130$dB, $1.055$dB and $1.036$dB, and their PVD values are 1.558dB, 1.109dB, and 1.038dB, respectively.  For SSIM, we notice similar trends. Fig.~\ref{fig:std} shows the PSNR curves of HEVC baseline, MFQE and NL-ConvLSTM on the sequence \emph{TunnelFlag}. In Fig.~\ref{fig:std}, comparing with MFQE, our method gets similar improvement on PQFs, but achieves much higher PSNR and SSIM improvement on non-PQFs. All these results show that our method performs more stably than the baseline and MFQE.

\subsection{Qualitative Comparison}
Fig.~\ref{fig:artifact} compares the reduction effect of different methods on four kinds of compression artifacts happened to images from Vimeo-90K with $qp$=37. The four artifacts are blocking, color bleeding, mosquito noise and ringing~\cite{zeng2014characterizing}. Each row stands for an image with a kind of compression artifacts. Concretely, in the 1st row, the cup edge is blurred due to blocking artifacts; and in the 2nd row, black color of hair overlapping the face results in a black stripe on the face. Words in the 3rd image are surrounded by mosquito noise, and the 4th image suffers from a silhouette-like shade parallel to the outline of the person. The last column shows the original frames, and the 8th column presents a cropped part from each original frame. The 1st column lists the compressed image of each cropped part. From the 2nd column to the 7th column, the cropped parts after artifact reduction by different methods are illustrated. Checking the images in Fig.~\ref{fig:artifact}, we can see that the cropped images after artifact reduction by our method (in the 7th column) have higher quality than these processed by the other methods, and are more similar to the original images (the 8th column). This means that our method can handle these distortions better than the five existing methods.

\subsection{Run Time Comparison}
In Tab.~\ref{run time} we present run-time comparison results. As our method has to process more frames, it consumes more time than the other methods. However, the run-time of our method is acceptable, and our method uses much less time than the original NL method (the NL-ConvLSTM without the approximation mechanism). Here, we give only the original NL's run-time at resolution 180$\times$180~(for higher resolutions, it consumes too much time and GPU-memory). Our method can be further sped up by cudnn-accelerated ConvLSTM and highly-tuned implementation.

\subsection{Applying to Other Compression Standard}
Finally, we check whether our method is effective for compressed video sequences generated by other compression algorithms, such as H.264.
The models are initialized with the corresponding models trained on video clips compressed by HEVC, and then are fine-tuned on video clips compressed by x264 in FFmpeg with $QP$=37.
The resulting PSNR/SSIM improvements of our method on Vimeo-90K and Yang \etal's dataset are 1.43dB/0.011 and 0.693dB/0.0085. The enhancement performance of our method on H.264 compressed videos is comparable to that on HEVC compressed videos. These results indicate that our method is also effective for H.264 compressed videos.

\section{Conclusion}
In this paper, we propose a novel end-to-end non-local ConvLSTM for video compression artifact removal by modeling spatial deformations and temporal variations simultaneously. 
Experiments on two datasets show that 
our method can enhance the quality of compressed videos considerably, remove artifacts effectively and outperform the existing  methods. 
In the future, we plan to extend our method to other low-level video enhancement tasks, such as video super-resolution, interpolation and frame prediction. 


{\small
\bibliographystyle{ieee_fullname}
\bibliography{egbib}
}

\end{document}